# On Remote Estimation with Multiple Communication Channels

Xiaobin Gao, Emrah Akyol, *Senior Member, IEEE,* and Tamer Başar, *Life Fellow, IEEE*

*Abstract*—This paper considers a sequential sensor scheduling and remote estimation problem with multiple communication channels. Departing from the classical remote estimation paradigm, which involves one communication channel (noiseless or noisy), we consider here the more realistic setting of two channels with different characteristics (one is cheap but noisy, the other one is costly but noiseless). We first show, via a counter-example, that the common folklore of applying symmetric threshold-based policy, which is well known to be optimal (for unimodal state densities) in the classical remote estimation problem, can no longer be optimal in our setting. In view of that, and in order to make the problem tractable, we introduce a side channel which signals to the receiver the sign of the underlying state. We show, under some technical assumptions, that a threshold-in-threshold based communication scheduling is optimal. The impact of the results is analyzed numerically based on dynamic programming. This numerical analysis reveals some rather surprising results inheriting known properties from the single channel settings, such as not exhausting all the opportunities available for the noisy channel.

## I. INTRODUCTION

Sensor scheduling and remote estimation problems arise in the applications of wireless sensor networks, which have been studied for decades and are still drawing extensive attention [2]–[18]. One setting where they arise is networked control systems, where one needs to collect information on the state of a remote plant to generate the control signal to be applied to the plant. To gather the information on the state, sensors are placed at the remote plant. The sensors take measurements on the state, and transmit their measurements to a receiver via wireless communication. Based on messages collected from the sensors, the receiver generates an estimate of the state, and the estimate is used in decision making, such as generating control signals. On the one hand, the quality of the estimate is crucial to decision making, and more transmissions from the sensors help improve the quality of the estimate. On the other hand, the sensors have limited energy for communication and

This research was supported in part by NSF under grant CCF 11-11342, in part by the U.S. Army Research Office (ARO) grant W911NF-16-1-0485, and in part by the Office of Naval Research (ONR) MURI grant N00014-16-1-2710. A paper including partial contents from this paper was presented at the 2016 American Control Conference [1].

X. Gao is with Coordinated Science Laboratory, University of Illinois at Urbana-Champaign, Urbana, IL 61801, email: xgao16@illinois.edu

E. Akyol is with Electrical and Computer Engineering Department, The State University of New York at Binghamton, Binghamton, NY, 13902, email: eakyol@binghamton.edu

T. Başar is with Coordinated Science Laboratory, University of Illinois at Urbana-Champaign, Urbana, IL 61801, email: basar1@illinois.edu

they are not able to make transmissions all the time throughout the time horizon of relevance. Therefore, there is need to devise a communication scheduling policy for the sensors so as to wisely allocate their transmission opportunities, and also devise an estimation policy for the receiver to best utilize the partial information received from the sensors, such that the cumulative estimation error over the time horizon is minimized subject to the communication constraints for the sensors.

This paper follows the line of research reported in [19]–[21]. In [19], the following problem was considered: estimate a one-dimensional Gauss-Markov process over a decision horizon of length $T$ using only $N \leq T$ measurements. Thus, over the decision horizon of length $T$, the sensor has exactly $N$ opportunities to transmit its observation to the estimator. This type of a communication constraint is called *hard constraint*. The sensor is restricted to apply *threshold-based policies*, that is, the sensor transmits its observation when the difference between the actual observation and the expected observation exceeds some threshold. With this assumption, it was shown in [19] that there exists a unique optimal threshold. In addition, the optimal estimation policy turns out to be Kalman filter-like. Later in [20], a similar problem was considered with, however, no hard constraint on transmission opportunities, but instead a cost associated with each transmission. This type of a communication constraint is called *soft constraint*. Using majorization theory and related techniques, it was shown in [20] that there exists a globally optimal communication scheduling policy within the class of threshold-based policies. With this result, there is no loss of optimality by restricting the sensor to apply threshold-based policies in [19]. The work in [21] extended the results of [19], [20] to a more general case where the sensor is equipped with an energy harvester, and the stochastic process has non-Gaussian but symmetric and unimodal distribution.

The communication between the sensor and the estimator has been assumed to be perfect in [19]–[21], which may not be realistic even though it was an important first step. Naturally, the next step was then to study the effect of imperfect communication, e.g., packet drop, delay, and channel noise on the sensor scheduling and remote estimation problems. In [22], [23], the settings with independent and identically distributed (i.i.d.) packet-dropping channel and Markov packet-dropping channel were considered. In [24], the problem with a random delay channel was studied. In [25], the problem with an additive noise channel was considered. Because of the presence of channel noise, the sensor needs to encode the message before transmission, and the estimator needs to decode the noise-corrupted message for estimation. Then, a zero-delay source-channel coding problem is involved, which introduces fundamental difficulties into the problem. By applying results discussed in [26], [27], it was shown in [25] that a threshold-based communication scheduling policy together with a pair of affine encoding/decoding policies are jointly optimal. In addition, it is worth noting that the sensor scheduling and remote estimation problems with adversary and/or power allocation during the transmission process draw increasing attentions recently. More details on these topics can be found in [28]–[30] and references therein.

In this paper, we have a combination of the perfect channel setting studied in [19], [20] and the noisy channel setting studied in [25], [31], [32]. For each observation, the sensor chooses among non-transmission, transmission over noisy channel, and transmission over perfect yet costly channel. In view of the results for the single channel (noisy or noiseless) setting, one may tend to conjecture that the optimal communication scheduling policy is threshold-in-threshold based. Specifically, the sensor computes the difference between its actual observation and the expected observation, and then compares the difference against the two thresholds. Depending on the value of the difference, that is, below the smaller threshold, between two thresholds, or above the larger threshold, the sensor will choose not to transmit, transmit over noisy channel, or transmit over perfect channel, correspondingly. Surprisingly, we show here via a counter example that this thinking does not hold, even though the stochastic process has symmetric and unimodal density. This renders the problem fairly difficult to solve. We first articulate the reason for this surprising result. Then, we explain how the presence of a side channel signaling the sign of the underlying state between the encoder and the decoder helps to facilitate the analysis. With this additional element along with some technical assumptions, we show the optimality of threshold-in-threshold based policy.[1] Armed with this result, we numerically obtain the optimal decision sequence, that is, the evolution of optimal thresholds in time. This numerical analysis demonstrates some rather surprising results inheriting known properties from the noisy and noiseless settings. For example, the sensor uses all communication opportunities for the perfect channel, yet may not use all communication opportunities for the noisy channel.

The rest of paper is organized as follows: in Section II, we formulate two optimization problems with soft and hard constraints separately. In Section III, we consider the optimization problem with soft constraint, whose results can be used to solve the optimization problem with hard constraint considered in Section IV. In Section V, we present numerical results for the problem with hard constraint. Finally in Section VI, we draw concluding remarks and identify future directions for research.

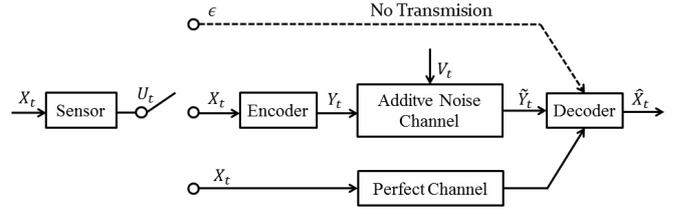

Fig. 1. System model

## II. PROBLEM FORMULATION

### A. System Model

Consider a discrete time communication scheduling and remote estimation problem over a finite time horizon, i.e., $t = 1, 2, \ldots, T$. A one-dimensional source process $\{X_t\}$ is an independent identically distributed (i.i.d.) stochastic process with probability density function $p_X$. At time $t$, the sensor observes the state of the source $X_t$. Then, it decides whether and how to transmit its observation to the remote estimator (which is also called "decoder"). Let $U_t \in \{0, 1, 2\}$ be the sensor's decision at time $t$. $U_t = 0$ means that the sensor chooses not to transmit its observation to the decoder, and hence it sends a free symbol $\epsilon$ to the decoder representing that nothing is transmitted. $U_t = 1$ means that the sensor chooses to transmit its observation to the decoder over an additive noise channel. Therefore, the sensor sends $X_t$ to an encoder, which then sends an encoded message, call it $Y_t$, to the communication channel. $Y_t$ is corrupted by an additive channel noise $V_t$. The noise process $\{V_t\}$ is a one-dimensional i.i.d. stochastic process with density $p_V$, which is independent of $\{X_t\}$. The encoder has average power constraint, that is,

$$\mathbb{E}[Y_t^2 | U_t = 1] \leq P_T,$$

where $P_T$ is known and constant for all $t$. When $U_t = 2$, sensor chooses to transmit its observation over a perfect channel. Hence, the decoder will receive $X_t$. Let $\tilde{Y}_t$ be the message received by decoder at time $t$; we have

$$\tilde{Y}_t = \begin{cases} \epsilon, & \text{if } U_t = 0 \\ Y_t + V_t, & \text{if } U_t = 1 \\ X_t, & \text{if } U_t = 2 \end{cases}$$

After receiving $\tilde{Y}_t$, the decoder generates an estimate on $X_t$, denoted by $\hat{X}_t$. The decoder is charged for squared distortion $(X_t - \hat{X}_t)^2$.

### B. Communication Constraints

We consider two separate optimization problems, corresponding to two kinds of communication constraints. In the

---

first scenario, at each time $t$, the sensor is charged for its decision, i.e., there is a cost function $c(U_t)$ such that

$$c(U_t) = \begin{cases} 0, & \text{if } U_t = 0 \\ c_1, & \text{if } U_t = 1 \\ c_2, & \text{if } U_t = 2 \end{cases}$$

Here, we have $c_2 > c_1 > 0$, which means that usage of the perfect channel is more costly than that of the noisy channel. $c_1, c_2$ are called the communication costs for using the noisy channel and the perfect channel, respectively. Such a communication constraint is called *soft constraint*. In the second scenario, the sensor is not charged for transmitting its observations, but is instead restricted to use the noisy channel and the perfect channel for no more than $N_1$ and $N_2$ times, respectively, i.e.,

$$\sum_{t=1}^{T} \mathbf{1}_{\{U_t=1\}} \leq N_1, \quad \sum_{t=1}^{T} \mathbf{1}_{\{U_t=2\}} \leq N_2,$$

where $\mathbf{1}_{\{\cdot\}}$ is the indicator function, and $N_1$, $N_2$ are positive integers. Such a communication constraint is called *hard constraint*.

### C. Decision Strategies

We make similar assumptions on the information structure to the ones made in [25, Section 2.3], yet apply them to a different problem.

We assume that at time $t$, the sensor has perfect recall of all its measurements by $t$, denoted by $X_{1:t}$, and of all the decisions it has made by $t-1$, denoted by $U_{1:t-1}$. The sensor makes decision $U_t$ based on its current information $(X_{1:t}, U_{1:t-1})$, that is,

$$U_t = f_t(X_{1:t}, U_{1:t-1}),$$

where $f_t$ is the sensor scheduling policy at time $t$ and $\mathbf{f} = \{f_1, f_2, \ldots, f_T\}$ is the sensor scheduling strategy.

We also assume that at time $t$, no matter whether and how the sensor decides to transmit the source output, it always transmits its decision $U_t$ to the encoder[2]. Let $\tilde{X}_t$ be the message received by the encoder at time $t$. Then,

$$\tilde{X}_t = \begin{cases} (X_t, U_t), & \text{if } U_t = 1 \\ U_t, & \text{otherwise} \end{cases}$$

Denote by $\tilde{X}_{1:t}$ the messages received by the encoder up to time $t$. Similar to the above, we assume that the encoder has perfect recall of $\tilde{X}_{1:t}$, and of all the encoded messages it has sent to the communication channel by $t-1$, denoted by $Y_{1:t-1}$[3]. The encoder generates the encoded message $Y_t$ based on its current information $(\tilde{X}_{1:t}, Y_{1:t-1})$, that is,

$$Y_t = g_t(\tilde{X}_{1:t}, Y_{1:t-1}),$$

where $g_t$ is the encoding policy at time $t$ and $\mathbf{g} = \{g_1, g_2, \ldots, g_T\}$ is the encoding strategy.

Finally, we assume that the decoder can deduce $U_t$ from $\tilde{Y}_t$. Furthermore, it is assumed that at time $t$, the decoder has perfect recall of all the messages received by $t$, denoted by $\tilde{Y}_{1:t}$, and of all the estimates it has generated by $t-1$, denoted by $\hat{X}_{1:t-1}$. The decoder generates the estimate $\hat{X}_t$ based on its current information $(\tilde{Y}_{1:t}, \hat{X}_{1:t-1})$, namely,

$$\hat{X}_t = h_t(\tilde{Y}_{1:t}, \hat{X}_{1:t-1}),$$

where $h_t$ is the decoding policy at time $t$ and $\mathbf{h} = \{h_1, h_2, \ldots, h_T\}$ is the decoding strategy.

In particular, we call the sensor, the encoder, and the decoder the *decision makers*. We call $(f_t, g_t, h_t)$ the *decision policies* at time $t$, and $(\mathbf{f}, \mathbf{g}, \mathbf{h})$ the *decision strategies*.

**Remark 1.** *At time $t$, the sensor's decisions by $t-1$, namely $U_{1:t-1}$, is a common information shared by all the decision makers. This is an important property, which will be used when solving the optimization problem under the hard constraint.*

### D. Optimization Problems

Consider the setting described above, with time horizon $[1, T]$, the probability density functions $p_X$ and $p_V$, and the power constraint $P_T$ as given.

Optimization problem with soft constraint: Given the communication cost function $c(\cdot)$, determine the decision strategies $(\mathbf{f}, \mathbf{g}, \mathbf{h})$ minimizing the cost functional

$$J(\mathbf{f}, \mathbf{g}, \mathbf{h}) := \mathbb{E}\left\{ \sum_{t=1}^{T} c(U_t) + (X_t - \hat{X}_t)^2 \right\}.$$

Optimization problem with hard constraint: Given the numbers of communication opportunities $N_1$ and $N_2$, as described earlier, and taking them as hard constraints not to be exceeded, determine the decision strategies $(\mathbf{f}, \mathbf{g}, \mathbf{h})$ minimizing the cost functional

$$J(\mathbf{f}, \mathbf{g}, \mathbf{h}) := \mathbb{E}\left\{ \sum_{t=1}^{T} (X_t - \hat{X}_t)^2 \right\}.$$

## III. Optimization Problem with Soft Constraint

### A. Counter intuitive property of the optimal communication scheduling policy

Since the source and the noise processes are i.i.d., by an argument similar to that in [25, Theorem 1], the optimization decision strategies can be obtained by solving a single-stage problem, as described in the following theorem.

**Theorem 1.** *Consider the optimization problem with soft constraint formulated in Section II-D. Without loss of optimality, the decision makers can apply decision policies restricted to the form*

$$U_t = f_t(X_t), \quad Y_t = g_t(\tilde{X}_t), \quad \hat{X}_t = h_t(\tilde{Y}_t), \quad t = 1, 2, \ldots, T,$$

*where $(f_t, g_t, h_t)$ are designed to minimize the instantaneous cost functional*

$$J_t(f_t, g_t, h_t) := \mathbb{E}[cU_t + (X_t - \hat{X}_t)^2].$$

---

[2]In practice, the sensor and the encoder are built together.

[3]If the sensor decides not to transmit its observation over the noisy channel, the encoder will not send anything to the noisy channel. In this scenario, we write $Y_t = \epsilon$.

*Furthermore,*

$$f_1 = f_2 = \ldots = f_T,$$
$$g_1 = g_2 = \ldots = g_T,$$
$$h_1 = h_2 = \ldots = h_T.$$

For simplicity, we henceforth suppress the subscript for time in this subsection. We further make the following assumptions on the optimization problem.

**Assumption 1.** *The source density $p_X$ is symmetric and unimodal around zero, i.e.,*

$$p_X(x) = p_X(-x), \ \forall \, x \in \mathbb{R}$$
$$p_X(x_1) \geq p_X(x_2), \ \forall \, |x_1| \geq |x_2|$$

**Assumption 2.** *The communication channel noise $V$ has zero mean and finite variance, denoted by $\sigma_V^2$.*

**Assumption 3.** *When the sensor decides to transmit its observation via the noisy channel, the encoder and decoder will apply affine policies in the form*

$$g(X) = \alpha\big(X - \mathbb{E}[X|U=1]\big)$$
$$h(\tilde{Y}) = \frac{1}{\alpha}\frac{\gamma}{\gamma+1}\tilde{Y} + \mathbb{E}[X|U=1]$$

*where $\gamma := P_T/\sigma_V^2$ is the signal-to-noise ratio (SNR), and $\alpha := \sqrt{P_T/\mathrm{Var}(X|U=1)}$ is the amplifying ratio. $\mathrm{Var}(X|U=1)$ is the variance of $X$ conditioned on the event that the sensor transmits the source output over the noisy channel.*

**Remark 2.** *The optimization problem (either with soft constraint or with hard constraint) can be viewed as concatenation of a communication scheduling problem and a zero-delay source-channel coding problem. For the zero-delay source-channel coding problem, it is well known that affine encoding and decoding policies are optimal if the source and noise have jointly Gaussian distribution. A more recent result (see [26]) states that the optimality of affine encoding/decoding policies still holds if the characteristic functions of source and noise satisfy a "matching condition". However, the source-channel coding problem in a general case is fairly difficult to solve, which is the case occurring here: due to the concatenation, the communication scheduling problem affects the source-channel coding problem by "reshaping" the source density. For example, consider the case where the source has Gaussian distribution and the sensor decides to transmit its observation if the observation lies outside some interval. Then, the density of the source, conditioned on the event that the sensor's observation lies outside some interval, will not be Gaussian anymore, which renders the problem generally intractable. Hence, we restrict here the encoder and the decoder to apply affine policies, and it is easy to see that the pair described above is optimal among the affine class.*

Note that the source density is symmetric around zero. Moreover, the distortion metric is the squared error, which is also symmetric around zero. It is intuitive to have a guess that the optimal communication scheduling policy is symmetric around zero. Also note that in an asymptotic case where the communication channel is noiseless, the optimal communication scheduling policy is symmetric around zero (as shown in [20]). Hence, we make the following assumption before proceeding further.

**Assumption 4.** *The sensor will apply a communication scheduling policy in the form*

$$f(x) = f(-x), \quad \forall \, x \in \mathbb{R}.$$

The following corollary to Theorem 1 is a consequence of Assumptions 1-4, which states that under these assumptions, the optimal communication scheduling policy is threshold-in-threshold based.

**Corollary 1.** *Consider the single-stage problem with Assumptions 1-4 holding. Then, the optimal communication scheduling policy is of the threshold-in-threshold type:*

$$f(x) = \begin{cases} 0, & \text{if } |x| \leq \beta_1 \\ 1, & \text{if } \beta_1 < |x| \leq \beta_2 \\ 2, & \text{if } |x| > \beta_2 \end{cases} \tag{1}$$

*where the parameters $\beta_1$ and $\beta_2$ are the "thresholds", with $0 < \beta_1 \leq \beta_2 < \infty$.*

Before proving Corollary 1, we first introduce some notations. Let $\mathcal{T}_0^f$, $\mathcal{T}_1^f$, $\mathcal{T}_2^f$ be the *non-transmission region*, the *noisy transmission region*, and the *perfect transmission region*, respectively, under the communication policy $f$, i.e.,

$$\mathcal{T}_i^f := \{x \in \mathbb{R}|f(x)=i\}, \ \ i \in \{0,1,2\}.$$

Consider the cost functional $J(f,g,h)$ associated with any group of decision policies $(f,g,h)$ satisfying Assumption 3[4] and any communication channel noise satisfying Assumption 2; then we have

$$J(f,g,h)$$
$$= \mathbb{E}\big[c(U) + (X - \hat{X})^2\big]$$
$$= \sum_{i \in \{0,1,2\}} \mathbb{E}\big[c(U) + (X - \hat{X})^2\big|X \in \mathcal{T}_i^f\big] \cdot \mathbb{P}(X \in \mathcal{T}_i^f).$$

We now have the following three properties:

(i) When $X \in \mathcal{T}_0^f$, the sensor decides not to transmit its observation. Then, when the optimal estimator is the conditional mean $\mathbb{E}[X|X \in \mathcal{T}_0^f]$. Moreover, we have

$$\mathbb{E}[(X - \hat{X})^2|X \in \mathcal{T}_0^f]$$
$$= \mathbb{E}\Big[\big(X - \mathbb{E}[X|X \in \mathcal{T}_0^f]\big)^2\big|X \in \mathcal{T}_0^f\Big]$$
$$= \mathrm{Var}(X|X \in \mathcal{T}_0^f).$$

---

[4]Here we do not place any restriction on $f$, which may or may not be symmetric around zero. When conducting the analysis, we consider the non-degenerate case where the probability measures over $\mathcal{T}_0^f$, $\mathcal{T}_1^f$, and $\mathcal{T}_2^f$ are non-zero. However, the analysis can easily adapt to the degenerate case.

(ii) When $X \in \mathcal{T}_1^f$, the sensor decides to transmit its observation over the noisy channel. By Assumptions 3, we have

$$
\begin{aligned}
\hat{X} &= \frac{1}{\alpha} \frac{\gamma}{\gamma+1} \tilde{Y} + \mathbb{E}[X|X \in \mathcal{T}_1^f] \\
&= \frac{1}{\alpha} \frac{\gamma}{\gamma+1} (Y+V) + \mathbb{E}[X|X \in \mathcal{T}_1^f] \\
&= \frac{1}{\alpha} \frac{\gamma}{\gamma+1} \Big( \alpha \big( X - \mathbb{E}[X|X \in \mathcal{T}_1^f] \big) + V \Big) \\
&\quad + \mathbb{E}[X|X \in \mathcal{T}_1^f] \\
&= \frac{\gamma}{\gamma+1} X + \frac{1}{\gamma+1} \mathbb{E}[X|X \in \mathcal{T}_1^f] + \frac{1}{\alpha} \frac{\gamma}{\gamma+1} V.
\end{aligned}
$$

Furthermore, the mean squared error conditioning on $X \in \mathcal{T}_1^f$ can be computed as

$$
\begin{aligned}
&\mathbb{E}[(X-\hat{X})^2 | X \in \mathcal{T}_1^f] \\
&= \mathbb{E}\Big[ \Big( X - \frac{\gamma}{\gamma+1} X - \frac{1}{\gamma+1} \mathbb{E}[X|X \in \mathcal{T}_1^f] \\
&\qquad - \frac{1}{\alpha} \frac{\gamma}{\gamma+1} V \Big)^2 | X \in \mathcal{T}_1^f \Big] \\
&= \frac{1}{(\gamma+1)^2} \mathbb{E}[(X - \mathbb{E}[X|X \in \mathcal{T}_1^f])^2 | X \in \mathcal{T}_1^f] \\
&\qquad + \frac{1}{\alpha^2} \frac{\gamma^2}{(\gamma+1)^2} \mathbb{E}[V^2] \\
&= \frac{1}{(\gamma+1)^2} \mathrm{Var}(X|X \in \mathcal{T}_1^f) + \frac{1}{\alpha^2} \frac{\gamma^2}{(\gamma+1)^2} \sigma_V^2 \\
&= \frac{1}{(\gamma+1)^2} \mathrm{Var}(X|X \in \mathcal{T}_1^f) \\
&\qquad + \frac{\gamma}{(\gamma+1)^2} \mathrm{Var}(X|X \in \mathcal{T}_1^f) \\
&= \frac{1}{\gamma+1} \mathrm{Var}(X|X \in \mathcal{T}_1^f).
\end{aligned}
\tag{2}
$$

The second equality holds since $X$ and $V$ are independent. The third equality holds dues to the fact that $\mathbb{E}[V] = 0$ (Assumption 2). The fourth equality holds by the expressions of $\gamma$ and $\alpha$ in Assumption 3.

(iii) When $X \in \mathcal{T}_2^f$, the sensor decides to transmit its observation over the perfect channel, and thus the decoder simply reports $\hat{X} = X$.

Combining the three cases together, we have

$$
\begin{aligned}
&J(f,g,h) \\
&= \mathrm{Var}(X|X \in \mathcal{T}_0^f) \mathbb{P}(X \in \mathcal{T}_0^f) + c_1 \mathbb{P}(X \in \mathcal{T}_1^f) \\
&\quad + \frac{1}{\gamma+1} \mathrm{Var}(X|X \in \mathcal{T}_1^f) \mathbb{P}(X \in \mathcal{T}_1^f) + c_2 \mathbb{P}(X \in \mathcal{T}_2^f)
\end{aligned}
\tag{3}
$$

With the notations and properties above, we are now in a position to prove Corollary 1.

**Proof of Corollary 1.** Assumption 4 states that $\mathcal{T}_0^f$, $\mathcal{T}_1^f$, $\mathcal{T}_2^f$ are symmetric around zero[5]. Combining Assumptions 1 and 4, it is easy to see that

$$
\mathbb{E}[X|X \in \mathcal{T}_0^f] = \mathbb{E}[X|X \in \mathcal{T}_1^f] = \mathbb{E}[X|X \in \mathcal{T}_2^f] = 0.
$$

Then, the expected cost $J(f,g,h)$ in (3) can be further expressed as

$$
\begin{aligned}
&J(f,g,h) \\
&= \mathbb{E}[X^2|X \in \mathcal{T}_0^f]\mathbb{P}(X \in \mathcal{T}_0^f) + c_1 \mathbb{P}(X \in \mathcal{T}_1^f) \\
&\quad + \frac{1}{\gamma+1} \mathbb{E}[X^2|X \in \mathcal{T}_1^f]\mathbb{P}(X \in \mathcal{T}_1^f) + c_2 \mathbb{P}(X \in \mathcal{T}_2^f) \\
&= \int_{x \in \mathcal{T}_0^f} x^2 p_X(x) dx + \int_{x \in \mathcal{T}_1^f} (c_1 + \frac{1}{\gamma+1} x^2) p_X(x) dx \\
&\quad + \int_{x \in \mathcal{T}_2^f} c_2 p_X(x) dx \\
&=: \int_{x \in \mathbb{R}} \tilde{J}(x, f(x)) p_X(x) dx,
\end{aligned}
$$

where

$$
\tilde{J}(x, f(x)) = \begin{cases} x^2, & \text{if } f(x) = 0 \\ c_1 + \dfrac{1}{\gamma+1} x^2, & \text{if } f(x) = 1 \\ c_2, & \text{if } f(x) = 2 \end{cases}
$$

Hence, $J(f,g,h)$ can be minimized by $f^*$ satisfying

$$
f^*(x) = \arg\min_{u \in \{0,1,2\}} \tilde{J}(x, u),
$$

and $(g^*, h^*)$ induced by $f^*$ according to Assumption 3. Since $\tilde{J}(x,0)$, $\tilde{J}(x,1)$, and $\tilde{J}(x,2)$ are symmetric around zero, we only need to consider the case when $x \geq 0$. Let $\beta_{01} = \sqrt{(\gamma+1)c_1/\gamma}$ and $\beta_{02} = \sqrt{c_2}$. Since $1/(\gamma+1) < 1$, it is easy to check that

$$
\begin{aligned}
\tilde{J}(x,0) &\leq \tilde{J}(x,1), \ x \in [0, \beta_{01}]; \\
\tilde{J}(x,0) &> \tilde{J}(x,1), \ x \in (\beta_{01}, \infty),
\end{aligned}
$$

and

$$
\begin{aligned}
\tilde{J}(x,0) &\leq \tilde{J}(x,2), \ x \in [0, \beta_{02}]; \\
\tilde{J}(x,0) &> \tilde{J}(x,2), \ x \in (\beta_{02}, \infty).
\end{aligned}
$$

Let $\beta_1 = \min\{\beta_{01}, \beta_{02}\}$, and we have

$$
\begin{aligned}
\tilde{J}(x,0) &\leq \min\{\tilde{J}(x,1), \tilde{J}(x,2)\}, \ x \in [0, \beta_1]; \\
\tilde{J}(x,0) &> \min\{\tilde{J}(x,1), \tilde{J}(x,2)\}, \ x \in (\beta_1, \infty).
\end{aligned}
$$

Hence, $f^*(x) = 0$ when $x \in [0, \beta_1]$. Furthermore, when $x \in (\beta_1, \infty)$, we only need to compare $\tilde{J}(x,1)$ with $\tilde{J}(x,2)$, and one of the following two cases occurs:

(i) $c_1 + \dfrac{1}{\gamma+1} \beta_1^2 > c_2$, and then

$$
\tilde{J}(x,1) > \tilde{J}(x,2), \ \forall x \in (\beta_1, \infty),
$$

---

[5]However, $\mathcal{T}_0^f$, $\mathcal{T}_1^f$, $\mathcal{T}_2^f$ may or may not be connected.

which implies that $f^*(x) = 2$ when $x \in (\beta_1, \infty)$. Hence, $f^*$ is of the threshold-in-threshold type described by (1), with parameters $\beta_1 = \beta_2$.

(ii) $c_1 + \dfrac{1}{\gamma + 1}\beta_1^2 \le c_2$. Let $\beta_2 = \sqrt{(c_2 - c_1)(\gamma + 1)}$. It can be checked that

$$\tilde{J}(x, 1) \le \tilde{J}(x, 2), \ x \in (\beta_1, \beta_2];$$
$$\tilde{J}(x, 1) > \tilde{J}(x, 2), \ x \in (\beta_2, \infty).$$

Hence, $f^*(x) = 1$ when $x \in (\beta_1, \beta_2]$, and $f^*(x) = 2$ when $x \in (\beta_2, \infty)$. $f^*$ is of the threshold-in-threshold type. $\square$

Although Assumption 4 and Corollary 1 seem very intuitive at first glance, the following counter example renders them not valid from the point of global optimality.

*Counter example*: Consider the case where $X$ has uniform distribution over $[-L, L]$, namely,

$$p_X(x) = \frac{1}{2L}, \ x \in [-L, L].$$

Suppose that the parameters satisfy

$$\frac{\gamma + 1}{\gamma}c_1 < c_2; \quad \sqrt{(c_2 - c_1)(\gamma + 1)} < L. \quad (4)$$

By Corollary 1, the single-stage problem admits a solution including a symmetric communication scheduling policy $f^*$ of threshold-in-threshold type with parameters $\beta_1, \beta_2$, and a pair of encoding/decoding policies $(g^*, h^*)$ induced by $f^*$ according to Assumption 3. By (4), we have $0 < \beta_1 < \beta_2 < L$. Hence, the non-transmission region, the noisy transmission region, and the perfect transmission region corresponding to $f^*$ are as follows:

$$\mathcal{T}_0^{f^*} = [-\beta_1, \beta_1],$$
$$\mathcal{T}_1^{f^*} = [-\beta_2, -\beta_1) \cup (\beta_1, \beta_2],$$
$$\mathcal{T}_2^{f^*} = [-L, -\beta_2) \cup (\beta_2, L].$$

We now construct a non-symmetric communication scheduling policy $f'$ by specifying its non-transmission region, noisy transmission region, and perfect transmission region:

$$\mathcal{T}_0^{f'} = \mathcal{T}_0^{f^*},$$
$$\mathcal{T}_1^{f'} = (\beta_1, 2\beta_2 - \beta_1],$$
$$\mathcal{T}_2^{f'} = [-L, -\beta_1) \cup (2\beta_2 - \beta_1, L].$$

Since the source is uniformly distributed, we have

$$\mathbb{P}(X \in \mathcal{T}_1^{f'}) = \mathbb{P}(X \in \mathcal{T}_1^{f^*}) = \frac{\beta_2 - \beta_1}{L}.$$

Essentially, we rearrange the noisy transmission region, without changing its probability measure, to make the region connected. This procedure is illustrated in Fig. 2. Induced by $f'$, we obtain the encoding and decoding policies $(g', h')$ satisfying Assumption 3. Furthermore, by (3), we have

$$J(f', g', h') - J(f^*, g^*, h^*)$$
$$= \frac{\mathbb{P}(X \in \mathcal{T}_1^{f'})}{\gamma + 1}\Big(\text{Var}(X | X \in \mathcal{T}_1^{f'}) - \text{Var}(X | X \in \mathcal{T}_1^{f^*})\Big).$$

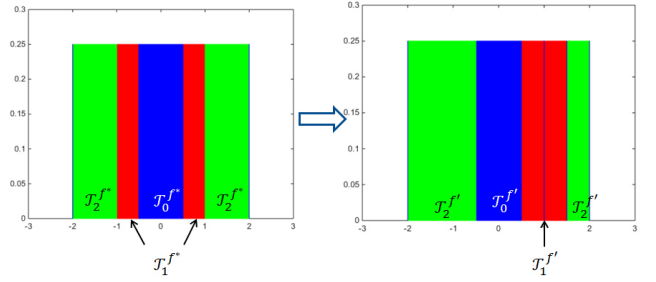

Fig. 2. The counter example

The regions $\mathcal{T}_1^{f'}$ and $\mathcal{T}_1^{f^*}$ have the same probability measure under uniform distribution, while $\mathcal{T}_1^{f'}$ is connected. Evidently, we have $\text{Var}(X | X \in \mathcal{T}_1^{f'}) < \text{Var}(X | X \in \mathcal{T}_1^{f^*})$, which implies $J(f') < J(f^*)$. Hence, the symmetric communication scheduling policy $f^*$ together with the encoding/decoding policies $(g^*, h^*)$ are not globally optimal.

**Remark 3.** *The counter example above uncovers a rather surprising result, namely, that with the presence of a noisy channel, the common folklore that the optimal communication scheduling policy is symmetric does not hold. As illustrated in the example, the noisy transmission region under symmetric communication policy is disconnected, which results in large conditional variance. Therefore, symmetric communication policy does not take full advantage of the presence of the noisy channel.*

The non-symmetric property of the optimal communication scheduling policy makes the problem fairly difficult to solve. In order to fix this issue and render the problem tractable, we further assume the existence of a side channel.

### B. Modified problem

We now assume that there exists a side channel between the encoder and the decoder. Recall that at time $t$, if the sensor decides to transmit its observation $X_t$ via the noisy channel, it sends the observation to the encoder. Then, the encoder sends an encoded message $Y_t$ to the noisy channel. We now assume that the encoder additionally sends the sign of $X_t$, denoted by $S_t$, to the decoder over the side channel, which is illustrated in Fig. 3. Assume that the side channel is noise-free[6]. Let $S_{1:t}$ be the collections of side-channel messages up to $t$. Now, the information available to the encoder and the decoder at time $t$ is $(\tilde{X}_{1:t}, S_{1:t}, Y_{1:t-1})$ and $(\tilde{Y}_{1:t}, S_{1:t}, \hat{X}_{1:t-1})$, respectively. The encoder and the decoder generate the encoded message $Y_t$ and estimate $\hat{X}_t$, respectively, according to

$$Y_t = g_t(\tilde{X}_{1:t}, S_{1:t}, Y_{1:t-1}), \quad \hat{X}_t = h_t(\tilde{Y}_{1:t}, S_{1:t}, \hat{X}_{1:t-1}).$$

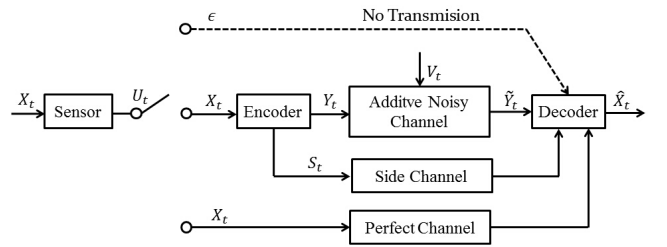

Fig. 3. Modified system

---

[6] A side-channel message $S_t \in \{-1, 1\}$ takes only one bit, and hence it can be sent reliably. When the sensor decides not to transmit its observation via the noisy channel, the encoder will not send anything to the side channel. In this case, we write $S_t = \epsilon$.

Similar to Theorem 1, it can be shown that without loss of optimality, the encoder and the decoder can ignore the information about the past states when making decisions, namely,

$$Y_t = g_t(\tilde{X}_t, S_t), \quad \hat{X}_t = h_t(\tilde{Y}_t, S_t).$$

Furthermore, the optimal decision strategies $(\mathbf{f}, \mathbf{g}, \mathbf{h})$ can be obtained by solving the single-stage problem, and hence we suppress the subscript for time in this subsection. It is important to note that the side channel enables the encoder and the decoder to apply different encoding and decoding policies for the positive and negative realizations of the source. Hence, we need to modify Assumption 3 (while keeping Assumptions 1 and 2).

**Assumption 5.** *When the sensor decides to transmit its observation over the noisy channel, the encoder and the decoder will apply piecewise affine policies, i.e.,*

$$\begin{aligned} g(X, S) &= S\alpha(S)\big(X - \mathbb{E}\left[X|U = 1, S\right]\big) \\ h(\tilde{Y}, S) &= \frac{1}{\alpha(S)}\frac{\gamma}{\gamma + 1}S\tilde{Y} + \mathbb{E}\left[X|U = 1, S\right] \end{aligned}$$

*The parameter $\gamma = P_T/\sigma_V^2$ is the signal-to-noise ratio, and $\alpha(S) = \sqrt{P_T/\mathrm{Var}(X|U = 1, S)}$ is the amplifying ratio. $\mathbb{E}[X|U = 1, S]$ and $\mathrm{Var}(X|U = 1, S)$ are the conditional mean and variance, respectively.*

We now compute the expected cost $J(f, g, h)$ associated with any communication scheduling policy $f$[7] and the encoding/decoding policies $(g, h)$ induced by $f$ under Assumption 5. Let $\mathcal{T}_{1+}^f$, $\mathcal{T}_{1-}^f$ be the *positive noisy transmission region* and the *negative noisy transmission region*, respectively, according to $f$, i.e.,

$$\mathcal{T}_{1+}^f := \{x > 0 | f(x) = 1\}, \quad \mathcal{T}_{1-}^f := \{x < 0 | f(x) = 1\}.$$

Similar to (2) and (3), it can be computed that

$$\mathbb{E}[(X - \hat{X})^2 | X \in \mathcal{T}_{1+}^f] = \frac{1}{\gamma + 1}\mathrm{Var}(X | X \in \mathcal{T}_{1+}^f)$$

$$\mathbb{E}[(X - \hat{X})^2 | X \in \mathcal{T}_{1-}^f] = \frac{1}{\gamma + 1}\mathrm{Var}(X | X \in \mathcal{T}_{1-}^f)$$

---

[7]Again, the communication scheduling policy $f$ considered here may or may not be symmetric around zero. When conducting the analysis, we consider the non-degenerate case where the probability measures over all transmission regions are non-zero. The analysis can easily be adapted to the degenerate case.

Moreover,

$$\begin{aligned} &J(f, g, h) \\ &= \mathrm{Var}(X | X \in \mathcal{T}_0^f)\mathbb{P}(X \in \mathcal{T}_0^f) + c_2\mathbb{P}(X \in \mathcal{T}_2^f) \\ &\quad + \frac{1}{\gamma + 1}\mathrm{Var}(X | X \in \mathcal{T}_{1+}^f)\mathbb{P}(X \in \mathcal{T}_{1+}^f) + c_1\mathbb{P}(X \in \mathcal{T}_{1-}^f) \\ &\quad + \frac{1}{\gamma + 1}\mathrm{Var}(X | X \in \mathcal{T}_{1-}^f)\mathbb{P}(X \in \mathcal{T}_{1-}^f) + c_1\mathbb{P}(X \in \mathcal{T}_{1+}^f). \end{aligned} \tag{5}$$

We compare the cost functionals in the original problem and modified problem, which are described by (3) with (5), respectively. It appears that the conditional variance over the noisy transmission region $\mathrm{Var}(X | X \in \mathcal{T}_1^f)$ is replaced by two conditional variances over the positive and negative noisy transmission regions, that is, $\mathrm{Var}(X | X \in \mathcal{T}_{1+}^f)$ and $\mathrm{Var}(X | X \in \mathcal{T}_{1-}^f)$. As discussed in Remark 3, a symmetric communication scheduling policy $f$ has a disconnected noisy transmission region $\mathcal{T}_1^f$, which may lead to a large conditional variance $\mathrm{Var}(X | X \in \mathcal{T}_1^f)$. Hence, a symmetric communication scheduling policy cannot be optimal in the original problem. However, in the modified problem, a symmetric communication scheduling policy $f$ may have connected positive and negative noisy transmission regions $\mathcal{T}_{1+}^f$ and $\mathcal{T}_{1-}^f$, which would result in small conditional variances $\mathrm{Var}(X | X \in \mathcal{T}_{1+}^f)$ and $\mathrm{Var}(X | X \in \mathcal{T}_{1-}^f)$. Therefore, a symmetric communication scheduling policy can be globally optimal, and hence Assumption 4 is a reasonable assumption for the modified problem. We keep this assumption and further establish the optimality of threshold-in-threshold policy, as stated in the following theorem.

**Theorem 2.** *Consider the modified problem with Assumptions 1,2, 4 and 5 holding. Without loss of optimality, the sensor can apply communication scheduling policy of threshold-in-threshold type described by (1).*

To prove Theorem 2, we need the following proposition.

**Proposition 1.** *Let $p_X$ be the probability density function of the random variable $X$, which we take to be symmetric and unimodal around zero. Consider two open intervals $(\beta_1, \beta_2)$ and $(\beta_1', \beta_2')$ such that $0 \le \beta_1 \le \beta_1'$ and $\mathbb{P}(X \in (\beta_1, \beta_2)) = \mathbb{P}(X \in (\beta_1', \beta_2'))$. Then,*

$$\mathrm{Var}(X | X \in (\beta_1, \beta_2)) \le \mathrm{Var}(X | X \in (\beta_1', \beta_2')).$$

**PROOF.** Let $k := \mathbb{P}(X \in (\beta_1, \beta_2))$. Consider any open interval $(\eta_1, \eta_2), \eta_1 \ge 0$ such that $\mathbb{P}(X \in (\eta_1, \eta_2)) = k$. Since

$$\mathbb{P}(X \in (\eta_1, \eta_2)) = \int_{\eta_1}^{\eta_2} p_X(x)dx = k,$$

taking derivative with respect to $\eta_1$, we have

$$-p_X(\eta_1) + \frac{d\eta_2}{d\eta_1}\, p_X(\eta_2) = 0. \tag{6}$$

Now consider the partial derivative of $\mathrm{Var}(X|X \in (\eta_1, \eta_2)) \cdot \mathbb{P}(X \in (\eta_1, \eta_2))$ with respect to $\eta_1$:

$$\frac{\partial}{\partial \eta_1} \mathrm{Var}\big(X|X \in (\eta_1, \eta_2)\big) \int_{\eta_1}^{\eta_2} p_X(x) dx$$

$$= \frac{\partial}{\partial \eta_1} \left( \int_{\eta_1}^{\eta_2} x^2 p_X(x) dx - \frac{\left( \int_{\eta_1}^{\eta_2} x p_X(x) dx \right)^2}{\int_{\eta_1}^{\eta_2} p_X(x) dx} \right)$$

$$= -\eta_1^2 p_X(\eta_1)$$

$$+ \frac{2\eta_1 p_X(\eta_1) \int_{\eta_1}^{\eta_2} x p_X(x) dx \cdot \int_{\eta_1}^{\eta_2} p_X(x) dx}{\left( \int_{\eta_1}^{\eta_2} p_X(x) dx \right)^2}$$

$$- \frac{\left( \int_{\eta_1}^{\eta_2} x p_X(x) dx \right)^2 \cdot p_X(\eta_1)}{\left( \int_{\eta_1}^{\eta_2} p_X(x) dx \right)^2}$$

$$= -p_X(\eta_1) \cdot \big( \eta_1 - \mathbb{E}[X|X \in (\eta_1, \eta_2)] \big)^2. \tag{7}$$

Similarly, we have

$$\frac{\partial}{\partial \eta_2} \mathrm{Var}\big(X|X \in (\eta_1, \eta_2)\big) \int_{\eta_1}^{\eta_2} p_X(x) dx$$

$$= p_X(\eta_2) \cdot \big( \eta_2 - \mathbb{E}[X|X \in (\eta_1, \eta_2)] \big)^2. \tag{8}$$

Combining (6)-(8),

$$\frac{d}{d\eta_1} \mathrm{Var}(X|X \in (\eta_1, \eta_2)) \mathbb{P}(X \in (\eta_1, \eta_2))$$

$$= \frac{\partial}{\partial \eta_1} \mathrm{Var}(X|X \in (\eta_1, \eta_2)) \mathbb{P}(X \in (\eta_1, \eta_2))$$

$$+ \frac{d\eta_2}{d\eta_1} \frac{\partial}{\partial \eta_2} \mathrm{Var}(X|X \in (\eta_1, \eta_2)) \mathbb{P}(X \in (\eta_1, \eta_2))$$

$$= p_X(\eta_1) \Big( \big( \eta_2 - \mathbb{E}[X|X \in (\eta_1, \eta_2)] \big)^2$$

$$- \big( \eta_1 - \mathbb{E}[X|X \in (\eta_1, \eta_2)] \big)^2 \Big) \tag{9}$$

Since $p_X(x)$ is unimodal around zero and thus non-increasing when $x \geq 0$, it is easy to see that

$$\eta_2 - \mathbb{E}[X|X \in (\eta_1, \eta_2)] \geq \mathbb{E}[X|X \in (\eta_1, \eta_2)] - \eta_1. \tag{10}$$

Combining (9) and (10), we further have

$$\frac{d}{d\eta_1} \mathrm{Var}(X|X \in (\eta_1, \eta_2)) \mathbb{P}(X \in (\eta_1, \eta_2))$$

$$= k \cdot \frac{d}{d\eta_1} \mathrm{Var}(X|X \in (\eta_1, \eta_2))$$

$$\geq 0.$$

The inequality above implies that when shifting interval $(\eta_1, \eta_2)$ while preserving its probability measure, we have

$$\frac{d}{d\eta_1} \mathrm{Var}(X|X \in (\eta_1, \eta_2)) \geq 0.$$

Integrating both sides from $\beta_1$ to $\beta_1'$ and by comparison principle, we establish the desired inequality. □

We are now in a position to prove Theorem 2. The idea of the proof is as follows: given any symmetric communication scheduling policy $f$, we can construct another symmetric communication scheduling policy $\tilde{f}$ achieving no greater cost. Analysis on $\tilde{f}$ shows that it is either threshold-in-threshold based or "threshold-in-threshold-in-threshold" based. For the second case, we can further construct another communication scheduling policy $f'$ of threshold-in-threshold type, which achieves no greater cost.

**Proof of Theorem 2.** Let $(f, g, h)$ satisfying Assumptions 4 and 5 be given. Since $f(x)$ and $p_X(x)$ are symmetric around zero, we have

$$\mathbb{E}[X|X \in \mathcal{T}_0^f] = 0,$$
$$\mathbb{E}[X|X \in \mathcal{T}_{1+}^f] = -\mathbb{E}[X|X \in \mathcal{T}_{1-}^f] =: b. \tag{11}$$

Then, the cost functional described by (5) can be further expressed as

$$J(f, g, h)$$

$$= \mathrm{Var}(X|X \in \mathcal{T}_0^f) \mathbb{P}(X \in \mathcal{T}_0^f) + c_2 \mathbb{P}(X \in \mathcal{T}_2^f)$$

$$+ c_1 \mathbb{P}(X \in \mathcal{T}_{1+}^f) + c_1 \mathbb{P}(X \in \mathcal{T}_{1-}^f)$$

$$+ \frac{1}{\gamma + 1} \mathrm{Var}(X|X \in \mathcal{T}_{1+}^f) \mathbb{P}(X \in \mathcal{T}_{1+}^f)$$

$$+ \frac{1}{\gamma + 1} \mathrm{Var}(X|X \in \mathcal{T}_{1-}^f) \mathbb{P}(X \in \mathcal{T}_{1-}^f)$$

$$= \int_{x \in \mathcal{T}_0^f} x^2 p_X(x) dx + \int_{x \in \mathcal{T}_2^f} c_2 p_X(x) dx$$

$$+ \int_{x \in \mathcal{T}_{1+}^f} \big( c_1 + \frac{1}{\gamma + 1} (x - b)^2 \big) p_X(x) dx$$

$$+ \int_{x \in \mathcal{T}_{1-}^f} \big( c_1 + \frac{1}{\gamma + 1} (x + b)^2 \big) p_X(x) dx$$

$$= \int_{x \in \mathcal{T}_0^f} x^2 p_X(x) dx + \int_{x \in \mathcal{T}_2^f} c_2 p_X(x) dx$$

$$+ \int_{x \in \mathcal{T}_{1+}^f \cup \mathcal{T}_{1-}^f} \big( c_1 + \frac{1}{\gamma + 1} (|x| - b)^2 \big) p_X(x) dx$$

$$=: \int_{x \in \mathbb{R}} \tilde{J}(x, f(x)) p_X(x) dx,$$

where the third equality holds since $x \in \mathcal{T}_{1-}^f$ implies that $x < 0$ and $(x + b)^2 = (|x| - b)^2$. In addition, $\tilde{J}(x, u)$ is defined as follows:

$$\tilde{J}(x, u) = \begin{cases} x^2, & \text{if } u = 0 \\ c_1 + \dfrac{1}{\gamma + 1} (|x| - b)^2, & \text{if } u = 1 \\ c_2, & \text{if } u = 2 \end{cases}$$

We now construct a communication scheduling policy $\tilde{f}$ such that

$$\tilde{f}(x) = \underset{u \in \{0,1,2\}}{\arg\min} \ \tilde{J}(x,u), \ \forall \ x \in \mathbb{R}.$$

Since $\tilde{J}(x,u)$ is symmetric in $x$ around zero, for each fixed $u$, it is easy to see that $\tilde{f}(x)$ is also symmetric around zero. Denote by $\tilde{b} := \mathbb{E}[X | X \in \mathcal{T}_{1+}^{\tilde{f}}]$ the conditional mean of the event that $X \in \mathcal{T}_{1+}^{\tilde{f}}$. Then, by the symmetry property of $\tilde{f}$ and $p_X$, we have $\mathbb{E}[X | X \in \mathcal{T}_{1-}^{\tilde{f}}] = -\tilde{b}$. Moreover, let $(\tilde{g}, \tilde{h})$ be the encoding/decoding policies induced by $\tilde{f}$ by Assumption 5, we have

$$
\begin{aligned}
& J(f,g,h) \\
\geq \ & \int_{x \in \mathbb{R}} \tilde{J}(x, \tilde{f}(x)) p_X(x) dx \\
= \ & \int_{x \in \mathcal{T}_0^{\tilde{f}}} x^2 p_X(x) dx + \int_{x \in \mathcal{T}_2^{\tilde{f}}} c_2 p_X(x) dx \\
& + \int_{x \in \mathcal{T}_{1+}^{\tilde{f}}} \left( c_1 + \frac{1}{\gamma+1}(x-b)^2 \right) p_X(x) dx \\
& + \int_{x \in \mathcal{T}_{1-}^{\tilde{f}}} \left( c_1 + \frac{1}{\gamma+1}(x+b)^2 \right) p_X(x) dx \\
\geq \ & \int_{x \in \mathcal{T}_0^{\tilde{f}}} x^2 p_X(x) dx + \int_{x \in \mathcal{T}_2^{\tilde{f}}} c_2 p_X(x) dx \\
& + \int_{x \in \mathcal{T}_{1+}^{\tilde{f}}} \left( c_1 + \frac{1}{\gamma+1}(x-\tilde{b})^2 \right) p_X(x) dx \\
& + \int_{x \in \mathcal{T}_{1-}^{\tilde{f}}} \left( c_1 + \frac{1}{\gamma+1}(x+\tilde{b})^2 \right) p_X(x) dx \\
= \ & J(\tilde{f}, \tilde{g}, \tilde{h}).
\end{aligned}
$$

The first inequality holds due to the way that $\tilde{f}$ is constructed. The second inequality holds since

$$
\begin{aligned}
& \int_{x \in \mathcal{T}_{1+}^{\tilde{f}}} (x-b)^2 p_X(x) dx \\
= \ & \mathbb{E}\big[(X-b)^2 | X \in \mathcal{T}_{1+}^{\tilde{f}}\big] \mathbb{P}(X \in \mathcal{T}_{1+}^{\tilde{f}}) \\
\geq \ & \mathbb{E}\big[(X-\tilde{b})^2 | X \in \mathcal{T}_{1+}^{\tilde{f}}\big] \mathbb{P}(X \in \mathcal{T}_{1+}^{\tilde{f}}) \\
= \ & \int_{x \in \mathcal{T}_{1+}^{\tilde{f}}} (x-\tilde{b})^2 p_X(x) dx,
\end{aligned}
$$

and similarly

$$
\int_{x \in \mathcal{T}_{1-}^{\tilde{f}}} (x+b)^2 p_X(x) dx \geq \int_{x \in \mathcal{T}_{1-}^{\tilde{f}}} (x+\tilde{b})^2 p_X(x) dx.
$$

The two inequalities above hold since $\tilde{b}$ and $-\tilde{b}$ are the conditional means of the events that $X \in \mathcal{T}_{1+}^{\tilde{f}}$ and $X \in \mathcal{T}_{1-}^{\tilde{f}}$, respectively, and thus they achieve the minimum mean squared errors.

We now analyze the structure of $\tilde{f}$. Since $\tilde{f}$ is symmetric around zero, we only need to consider the case where $x \geq 0$. It is easy to check that there exists $\beta_1 > 0$ such that

$$\tilde{J}(x,0) \leq \min\{\tilde{J}(x,1), \tilde{J}(x,2)\}, \ x \in [0, \beta_1];$$

$$\tilde{J}(x,0) > \min\{\tilde{J}(x,1), \tilde{J}(x,2)\}, \ x \in (\beta_1, \infty).$$

Hence, $\tilde{f}(x) = 0$ when $x \in [0, \beta_1]$, and we only need to compare $\tilde{J}(x,1)$ with $\tilde{J}(x,2)$ when $x \in (\beta_1, \infty)$. Note that $\tilde{J}(x,1)$ is parabolic opening upward, and $\tilde{J}(x,2)$ is a constant. Hence, one of the following three cases occurs when $x \in (\beta_1, \infty)$:

(i) $\tilde{J}(x,1)$ and $\tilde{J}(x,2)$ do not intersect, which implies

$$\tilde{J}(x,1) > \tilde{J}(x,2), x \in (\beta_1, \infty).$$

Therefore, $\tilde{f}(x) = 2$ when $x \in (\beta_1, \infty)$, and $\tilde{f}$ is of threshold-in-threshold type with $\beta_1 = \beta_2$.

(ii) $\tilde{J}(x,1)$ and $\tilde{J}(x,2)$ intersect only once at $x = \beta_2$, and

$$\tilde{J}(x,1) \leq \tilde{J}(x,2), \ x \in (\beta_1, \beta_2];$$

$$\tilde{J}(x,1) > \tilde{J}(x,2), \ x \in (\beta_2, \infty).$$

Then, $\tilde{f}(x) = 1$ when $x \in (\beta_1, \beta_2]$ and $\tilde{f}(x) = 2$ when $x \in (\beta_2, \infty)$. Hence, $\tilde{f}$ is of threshold-in-threshold type.

(iii) $\tilde{J}(x,1)$ and $\tilde{J}(x,2)$ intersect twice at $\beta_l$ and $\beta_r$, which implies

$$\tilde{J}(x,1) \leq \tilde{J}(x,2), \ x \in [\beta_l, \beta_r];$$

$$\tilde{J}(x,1) > \tilde{J}(x,2), \ x \in (\beta_1, \beta_l) \cup (\beta_r, \infty).$$

Hence, $\tilde{f}(x) = 1$ when $x \in [\beta_l, \beta_r]$ and $\tilde{f}(x) = 2$ when $x \in (\beta_1, \beta_l) \cup (\beta_r, \infty)$. Although $\tilde{f}$ is not in threshold-in-threshold form, yet we can construct a policy $f'$ of threshold-in-threshold type based on $\tilde{f}$, which achieves no greater cost. Let $f'$ be as follows:

$$
\begin{aligned}
\mathcal{T}_0^{f'} &= [-\beta_1, \beta_1], \\
\mathcal{T}_{1+}^{f'} &= (\beta_1, \beta_2], \quad \mathcal{T}_{1-}^{f'} = [-\beta_2, -\beta_1), \\
\mathcal{T}_2^{f'} &= (-\infty, -\beta_2) \cup (\beta_2, \infty),
\end{aligned}
$$

where $\beta_2$ is selected such that

$$\int_{\beta_1}^{\beta_2} p_X(x) dx = \int_{\beta_l}^{\beta_r} p_X(x) dx.$$

As illustrated in Fig. 4, we have shifted the positive and the negative transmission regions towards zero without changing the probability measure over the two regions. Let $(g', h')$ be the encoding and the decoding policies induced by $f'$ following Assumption 5. By (5), it can be computed that

$$
\begin{aligned}
& J(f', g', h') - J(\tilde{f}, \tilde{g}, \tilde{h}) \\
= \ & \frac{\mathbb{P}(X \in \mathcal{T}_{1+}^{\tilde{f}})}{\gamma+1} \big( \text{Var}(X | X \in \mathcal{T}_{1+}^{f'}) - \text{Var}(X | X \in \mathcal{T}_{1+}^{\tilde{f}}) \big) \\
& + \frac{\mathbb{P}(X \in \mathcal{T}_{1-}^{\tilde{f}})}{\gamma+1} \big( \text{Var}(X | X \in \mathcal{T}_{1-}^{f'}) - \text{Var}(X | X \in \mathcal{T}_{1-}^{\tilde{f}}) \big) \\
= \ & \frac{2\mathbb{P}(X \in \mathcal{T}_{1+}^{\tilde{f}})}{\gamma+1} \big( \text{Var}(X | X \in \mathcal{T}_{1+}^{f'}) - \text{Var}(X | X \in \mathcal{T}_{1+}^{\tilde{f}}) \big)
\end{aligned}
$$

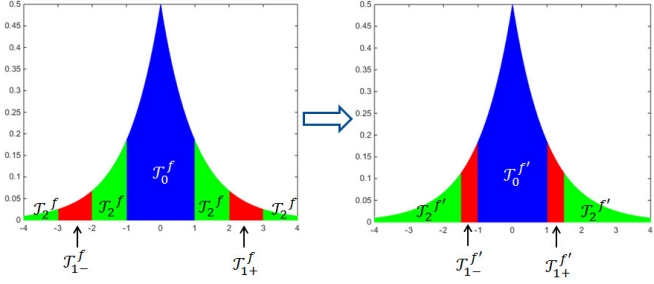

Fig. 4. Construction of $f'$ based on $\tilde{f}$

The second equality holds since $f'$ is symmetric around zero. Moreover, by Proposition 1, we have

$$\mathrm{Var}(X | X \in \mathcal{T}_{1+}^{f'}) \leq \mathrm{Var}(X | X \in \mathcal{T}_{1+}^{\tilde{f}}).$$

Hence, we conclude that

$$J(f', g', h') \leq J(\tilde{f}, \tilde{g}, \tilde{h}) \leq J(f, g, h),$$

and $f'$ is a communication scheduling policy of threshold-in-threshold type. $\square$

**Remark 4.** *The symmetry assumption on the communication scheduling policy, i.e., Assumption 4, is straightforward to understand and intuitive. Technically, this assumption can be relaxed when establishing the results stated in Theorem 2. In the proof of Theorem 2, Assumption 4 was used only to derive* (11), *which states that* $\mathbb{E}[X | X \in \mathcal{T}_{1+}^{f}] = -\mathbb{E}[X | X \in \mathcal{T}_{1-}^{f}]$. *On the other hand,* (11) *can be achieved not only by symmetric policies, but also by some non-symmetric policies. Hence, we can relax Assumption 4 to Assumption 6 (stated below), which directly states* (11) *as an assumption. We call Assumption 6 the "weak symmetry assumption".*

**Assumption 6.** *The sensor applies the communication scheduling policy $f$ such that*

*(i)* *If* $\mathbb{P}(X \in \mathcal{T}_0^f) > 0$, *then* $\mathbb{E}[X | X \in \mathcal{T}_0^f] = 0$.
*(ii)* *If* $\mathbb{P}(X \in \mathcal{T}_{1+}^f) > 0$ *and* $\mathbb{P}(X \in \mathcal{T}_{1-}^f) > 0$, *then* $\mathbb{E}[X | X \in \mathcal{T}_{1+}^f] = -\mathbb{E}[X | X \in \mathcal{T}_{1-}^f]$.

With Theorem 2, we have simplified an optimization problem over a function space to an optimization problem over a two-dimensional space. Hence, we can compute the optimal thresholds $\beta_1$ and $\beta_2$ via a standard approach. Once the communication scheduling policy $f$ is of threshold-in-threshold type with thresholds $\beta_1$ and $\beta_2$, the cost functional described by (5) can be further expressed as

$$
\begin{aligned}
& J(f, g, h) \\
={}& 2\int_0^{\beta_1} x^2 p_X(x)dx + 2c_2 \int_{\beta_2}^{\infty} p_X(x)dx \\
& + 2c_1 \int_{\beta_1}^{\beta_2} p_X(x)dx \\
& + \frac{2}{\gamma + 1} \mathrm{Var}\big(X | X \in (\beta_1, \beta_2)\big) \int_{\beta_1}^{\beta_2} p_X(x)dx.
\end{aligned}
$$

Taking partial derivative of $J(f, g, h)$ with respect to $\beta_1$, we have

$$
\begin{aligned}
& \frac{\partial J(f, g, h)}{\partial \beta_1} \\
={}& 2\beta_1^2 p_X(\beta_1) - 2c_1 p_X(\beta_1) \\
& + \frac{2}{\gamma + 1} \frac{\partial}{\partial \beta_1}\Big(\mathrm{Var}(X | X \in (\beta_1, \beta_2))\Big) \int_{\beta_1}^{\beta_2} p_X(x)dx
\end{aligned}
$$

Similar to the derivation in (7), it can be checked that

$$
\begin{aligned}
& \frac{\partial}{\partial \beta_1}\Big(\mathrm{Var}(X | X \in (\beta_1, \beta_2))\Big) \int_{\beta_1}^{\beta_2} p_X(x)dx \\
={}& -p_X(\beta_1)\Big(\beta_1 - \mathbb{E}[X | X \in (\beta_1, \beta_2)]\Big)^2.
\end{aligned}
$$

Hence, we have

$$
\begin{aligned}
& \frac{\partial J(f, g, h)}{\partial \beta_1} \\
={}& 2p_X(\beta_1)\left(\beta_1^2 - \frac{1}{\gamma + 1}\big(\beta_1 - \mathbb{E}[X | X \in (\beta_1, \beta_2)]\big)^2 - c_1\right)
\end{aligned}
\tag{12}
$$

We can also compute the partial derivative of $J(f, g, h)$ with respect to $\beta_2$ as follows:

$$
\begin{aligned}
& \frac{\partial J(f, g, h)}{\partial \beta_2} \\
={}& 2p_X(\beta_2)\left(\frac{1}{\gamma + 1}\big(\beta_2 - \mathbb{E}[X | X \in (\beta_1, \beta_2)]\big)^2 + c_1 - c_2\right)
\end{aligned}
\tag{13}
$$

By the first order optimality condition, the optimal thresholds should satisfy

$$
\begin{aligned}
& \beta_1^2 - \frac{1}{\gamma + 1}\big(\beta_1 - \mathbb{E}[X | X \in (\beta_1, \beta_2)]\big)^2 - c_1 = 0 \\
& \frac{1}{\gamma + 1}\big(\beta_2 - \mathbb{E}[X | X \in (\beta_1, \beta_2)]\big)^2 + c_1 - c_2 = 0
\end{aligned}
\tag{14}
$$

The existence and uniqueness of solution to (14) are difficult to analyze for general symmetric and unimodal densities. The reason is that $\mathbb{E}[X | X \in (\beta_1, \beta_2)]$ depends on the source density $p_X$, which might be complex. To simplify the analysis, we specify the source to have Laplace distribution with parameters $(0, \lambda^{-1})$, namely,

$$
p_X(x) = \begin{cases} \dfrac{1}{2}\lambda e^{-\lambda x}, & x \geq 0 \\ \dfrac{1}{2}\lambda e^{\lambda x}, & x < 0 \end{cases}
$$

Then, it can be computed that

$$
\begin{aligned}
\mathbb{E}[X | X \in (\beta_1, \beta_2)] &= \frac{1}{\lambda} + \beta_1 + \frac{(\beta_2 - \beta_1)e^{-\lambda(\beta_2 - \beta_1)}}{e^{-\lambda(\beta_2 - \beta_1)} - 1} \\
&=: \frac{1}{\lambda} + \beta_1 + \frac{\Delta\beta}{1 - e^{\lambda\Delta\beta}},
\end{aligned}
\tag{15}
$$

where $\Delta\beta = \beta_2 - \beta_1$. Plugging (15) into (14), we have

$$\beta_1 = \sqrt{c_1 + \frac{1}{\gamma+1}\left(\frac{1}{\lambda} + \frac{\Delta\beta}{1 - e^{\lambda\Delta\beta}}\right)^2},$$

$$\beta_2 - \beta_1 - \frac{\Delta\beta}{1 - e^{\lambda\Delta\beta}} = \frac{1}{\lambda} + \sqrt{(\gamma+1)(c_2 - c_1)},$$

which can be further simplified to

$$\beta_1 = \sqrt{c_1 + \frac{1}{\gamma+1}\left(\Delta\beta - \sqrt{(c_2 - c_1)(1+\gamma)}\right)^2},$$

$$\frac{\Delta\beta e^{\lambda\Delta\beta}}{e^{\lambda\Delta\beta} - 1} = \frac{1}{\lambda} + \sqrt{(c_2 - c_1)(1+\gamma)}.$$

(16)

Define a function $\varphi(x)$ in terms of $x$ as follows

$$\varphi(x) := \frac{xe^{\lambda x}}{e^{\lambda x} - 1} = \frac{x}{1 - e^{-\lambda x}}, \quad \forall x \in (0, \infty).$$

Then,

$$\frac{d\varphi(x)}{dx} = \frac{1 - e^{-\lambda x}}{(1 - e^{-\lambda x})^2} = \frac{1}{1 - e^{-\lambda x}} > 0, \quad \forall x \in (0, \infty),$$

which implies that $\varphi(x)$ is monotone increasing over $(0, \infty)$. Furthermore, it can be verified that $\varphi(x)$ ranges over $(1/\lambda, \infty)$ when $x \in (0, \infty)$. Hence, the second equation in (16) has a unique solution, which uniquely determines $\beta_1$ by the first equation in (16), and $\beta_2 = \Delta\beta + \beta_1$.

We denote by $(\beta_1^*, \beta_2^*)$ the unique solution to (16). We now show that $J(f, g, h)$ attains global minimum at $(\beta_1^*, \beta_2^*)$ among all pairs of thresholds $(\beta_1, \beta_2)$ satisfying $\beta_1 \le \beta_2$. We first fix $\beta_1$ and minimize $J(f, g, h)$ over $\beta_2 \in [\beta_1, \infty)$. It can be shown by analyzing $\partial J(f, g, h)/\partial\beta_2$ described in (13), that the minimizing $\beta_2$ is $\beta_2 = \beta_1 + \Delta\beta$, where $\Delta\beta$ satisfies the second equation in (16). Then, we keep $\beta_2$ as $\beta_2 = \beta_1 + \Delta\beta$, and minimize $J(f, g, h)$ over $\beta_1 \in [0, \infty)$. Taking the derivative of $J(f, g, h)$ with respect to $\beta_1$, we have

$$\frac{dJ(f, g, h)}{d\beta_1} = \frac{\partial J(f, g, h)}{\partial\beta_1} + \frac{\partial J(f, g, h)}{\partial\beta_2}\frac{d\beta_2}{d\beta_1} = \frac{\partial J(f, g, h)}{\partial\beta_1},$$

where the second equality holds since $\partial J(f, g, h)/\partial\beta_2 = 0$ when $\beta_2 = \beta_1 + \Delta\beta$. By analyzing $\partial J(f, g, h)/\partial\beta_1$ described in (12), it can be shown that the minimizing $\beta_1$ is the one satisfying the first equation in (16), which completes our argument.

**Remark 5.** *When formulating the problem in Section II, we had assumed that $c_1 < c_2$. If $c_1 \ge c_2$, that is, the noisy channel is more costly than the perfect channel, then the sensor should always use the perfect channel if it decides to transmit its observation. Therefore, the problem collapses to an optimization problem with one perfect channel. By the results from [20], the optimal communication scheduling policy is then still threshold-in-threshold type, with optimal thresholds $\beta_1 = \beta_2 = \sqrt{c_2}$.*

## IV. OPTIMIZATION PROBLEM WITH HARD CONSTRAINT

In this section, we still focus on the modified problem, but with hard constraint. We first introduce $E_t^n$ and $E_t^p$ as the remaining communication opportunities at time $t$ for the noisy channel and the perfect channel, respectively. Then, $E_t^n$ and $E_t^p$ can be obtained from the sensor's decisions up to $t - 1$, namely,

$$E_t^n = N_1 - \sum_{i=1}^{t-1}\mathbf{1}_{\{U_i=1\}}, \quad E_t^p = N_2 - \sum_{i=1}^{t-1}\mathbf{1}_{\{U_i=2\}}.$$

As discussed in Remark 1, $U_{1:t-1}$ is a common information shared by all the decision makers. Hence, $E_t^n$ and $E_t^p$ are also known by the sensor, the encoder, and the decoder. With a little abuse of the notation, we introduce $J^*(t, E_t^n, E_t^p)$ as the optimal cost-to-go when the system is initialized at time $t$ with $E_t^n$ and $E_t^p$ communication opportunities for the noisy channel and the perfect channel, respectively. Then, we have the following theorem on the structure of the optimal decision policies. Its proof is similar to that of Theorem 1 and hence is not included here.

**Theorem 3.** *Without loss of optimality, the sensor, the encoder, and the decoder can apply the following types of decision policies:*

$$U_t = f_t(X_t, E_t^n, E_t^p),$$

$$Y_t = g_t(X_t, S_t, E_t^n, E_t^p),$$

$$\hat{X}_t = h_t(\tilde{Y}_t, S_t, E_t^n, E_t^p).$$

*Furthermore, the optimal cost-to-go $J^*(t, E_t^n, E_t^p)$ can be obtained from the dynamic programming (DP) equation:*

$$J^*(t, E_t^n, E_t^p)$$

$$= \inf_{f_t, g_t, h_t}\left\{\mathbb{E}\big[(X_t - \hat{X}_t)^2 + J^*(t+1, E_{t+1}^n, E_{t+1}^p)\big]\right\}$$

*with the boundary condition $J^*(T+1, \cdot, \cdot) = 0$.*

Depending on the realization of $X_t$, $E_{t+1}^n$ may be $E_t^n$ or $E_t^n - 1$, and $E_{t+1}^p$ may be $E_t^p$ or $E_t^p - 1$. Hence, the DP equation can be written as

$$J^*(t, E_t^n, E_t^p)$$

$$= \inf_{f_t, g_t, h_t}\left\{\mathbb{E}\big[(X_t - \hat{X}_t)^2 + J^*(t+1, E_{t+1}^n, E_{t+1}^p)\big]\right\}$$

$$= J^*(t+1, E_t^n, E_t^p) + \inf_{f_t, g_t, h_t}\left\{\mathbb{E}\big[(X_t - \hat{X}_t)^2 + c_1(t, E_t^n, E_t^p)\mathbf{1}_{\{U_t=1\}} + c_2(t, E_t^n, E_t^p)\mathbf{1}_{\{U_t=2\}}\big]\right\}$$

$$= J^*(t+1, E_t^n, E_t^p) + \inf_{f_t, g_t, h_t}\left\{\mathbb{E}\big[(X_t - \hat{X}_t)^2 + c(t, E_t^n, E_t^p, U_t)\big]\right\}.$$

where

$$c_1(t, E_t^n, E_t^p) = J^*(t+1, E_t^n - 1, E_t^p) - J^*(t+1, E_t^n, E_t^p),$$

$$c_2(t, E_t^n, E_t^p) = J^*(t+1, E_t^n, E_t^p - 1) - J^*(t+1, E_t^n, E_t^p),$$

and

$$c(t, E_t^n, E_t^p, U_t) = \begin{cases} 0, & \text{if } U_t = 0 \\ c_1(t, E_t^n, E_t^p), & \text{if } U_t = 1 \\ c_2(t, E_t^n, E_t^p), & \text{if } U_t = 2 \end{cases}$$

Then the problem inside $\inf\{\cdot\}$ is a single-stage problem with soft constraint. Hence, we make the assumptions analogous to those we have made in Section III-B.

**Assumption 7.** *The source has Laplace distribution with parameters $(0, \lambda^{-1})$. The noise has zero mean and finite variance $\sigma_V^2$.*

**Assumption 8.** *The sensor will apply the communication scheduling policy $f_t$ such that*

*(i) If $\mathbb{P}\big(f_t(X_t, \cdot, \cdot) = 0\big) > 0$, then*
$$\mathbb{E}\big[X_t | f_t(X_t, \cdot, \cdot) = 0\big] = 0.$$

*(ii) If $\mathbb{P}\big(f_t(X_t, \cdot, \cdot) = 1, X_t > 0\big) > 0$ and $\mathbb{P}\big(f_t(X_t, \cdot, \cdot) = 1, X_t \leq 0\big) > 0$, then*
$$\mathbb{E}\big[X_t | f_t(X_t, \cdot, \cdot) = 1, X_t > 0\big]$$
$$= -\mathbb{E}[X_t | f_t(X_t, \cdot, \cdot) = 1, X_t \leq 0].$$

**Assumption 9.** *The encoder and the decoder are restricted to apply piecewise affine encoding and decoding policies, respectively, i.e.,*

$$g_t(X_t, S_t, E_t^n, E_t^p)$$
$$= S_t \cdot \alpha_t \cdot \big(X_t - \mathbb{E}\left[X_t | f_t(X_t, E_t^n, E_t^p) = 1, S_t\right]\big)$$

$$h_t(\tilde{Y}_t, S_t, E_t^n, E_t^p)$$
$$= \frac{1}{\alpha_t}\frac{\gamma}{\gamma+1} S_t \tilde{Y}_t + \mathbb{E}\left[X_t | f_t(X_t, E_t^n, E_t^p) = 1, S_t\right]$$

*where*

$$\gamma = \frac{P_T}{\sigma_V^2}, \quad \alpha_t = \sqrt{\frac{P_T}{\mathrm{Var}(X_t | f_t(X_t, E_t^n, E_t^p) = 1, S_t)}}$$

Then, we have the following theorem by applying the result of Theorem 2.

**Theorem 4.** *For the modified problem with Assumptions 7-9, the optimal communication scheduling policy is of threshold-in-threshold type as follows:*

$$f_t(X_t, E_t^n, E_t^p)$$
$$= \begin{cases} 0, & if \quad |X_t| \leq \beta_1(t, E_t^n, E_t^p) \\ 1, & if \quad \beta_1(t, E_t^n, E_t^p) < |X_t| \leq \beta_2(t, E_t^n, E_t^p) \\ 2, & if \quad |X_t| > \beta_2(t, E_t^n, E_t^p) \end{cases}$$

*where the optimal thresholds $\beta_1(t, E_t^n, E_t^p)$ and $\beta_2(t, E_t^n, E_t^p)$ can be obtained from (16) if $c_2(t, E_t^n, E_t^p) > c_1(t, E_t^n, E_t^p)$. Otherwise, both $\beta_1(t, E_t^n, E_t^p)$ and $\beta_2(t, E_t^n, E_t^p)$ are equal to $\sqrt{c_2(t, E_t^n, E_t^p)}$.*

## V. NUMERICAL RESULTS

In order to investigate the performance of the proposed decision strategies, we solve the DP equation numerically with $\lambda = 1$, $\gamma = 1$ and $T = 100$. We plot the optimal 100-stage estimation error versus the numbers of communication opportunities for the perfect channel and the noisy channel separately in two figures. We also generate a sample path of the numbers of remaining communication opportunities, $E_t^n$ and $E_t^p$, versus time. The numerical results have properties inheriting from both the setting with one perfect channel and the setting with one additive noise channel.

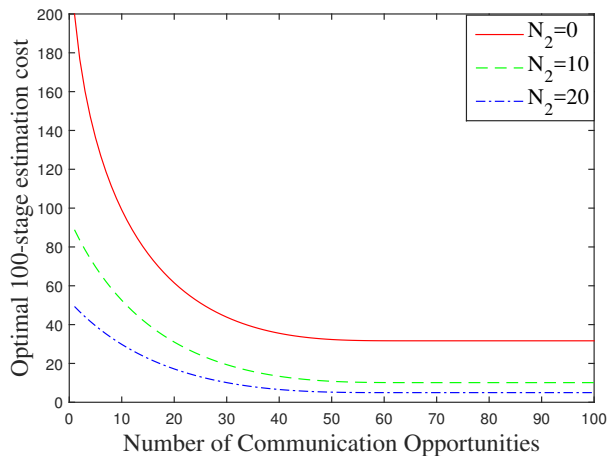

Fig. 5. Optimal 100-stage estimation error vs. number of communication opportunities for noisy channel

In Fig. 5, we fix the number of communication opportunities for perfect channel, $N_2 = 0, 10, 20$, respectively, and we plot the optimal 100-stage estimation error versus the number of communication opportunities for the noisy channel $N_1$. When $N_2 = 0$, there is no communication opportunity for the perfect channel, the problem collapses to the setting with one additive noise channel, which has been studied in [25]. As shown in [25, Fig.2], there exists an opportunity threshold such that the optimal 100-stage estimation error decreases when the number of communication opportunities is below the threshold, and remains constant above the threshold. The existence of opportunity threshold remains in the multi-channel setting. The reason for this interesting phenomenon has already been discussed in [25], but we recall it here also for completeness: since the sensor applies threshold-in-threshold based communication scheduling policy, an upper bound can be derived for the expected usage of the noisy channel, i.e., there exists $\bar{N}_1 < T$ such that

$$\sum_{t=1}^{T} \mathbb{E}[\mathbf{1}_{\{U_t = 1\}}] \leq \bar{N}_1.$$

If the number of communication opportunities for the noisy channel exceeds $\bar{N}_1$, then in the average sense, the sensor will not take advantage of the additional opportunities. Hence, the estimation error in the average sense, namely the mean squared error, will not further decrease.

Fig. 6 illustrates the performances of decision strategies when the number of communication opportunities for noisy channel, $N_1$, is fixed, and the number of communication opportunities for perfect channel, $N_2$, varies over $\{0, 1, \ldots, 100\}$. When $N_1 = 0$, there is no communication opportunity for the noisy channel, and hence the problem collapses to the one with one perfect channel, and the plot recovers the one in [19, Fig.5]. As shown in [19, Fig.5], the optimal 100-stage estimation error over the time horizon decreases to zero as the number of communication opportunities for the perfect channel increases to reach the length of time horizon. This trend remains for the multi-channel setting.

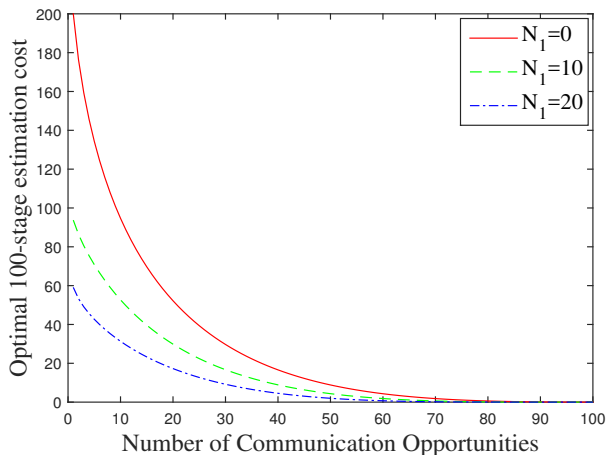

Fig. 6. Optimal 100-stage estimation error vs. number of communication opportunities for perfect channel

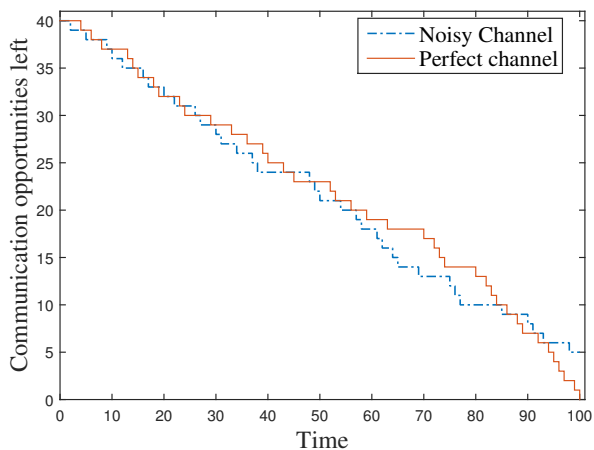

Fig. 7. Evolution of the remaining communication opportunities

Fig. 7 depicts a sample path illustrating the evolution of the remaining communication opportunities for the noisy and perfect channels, i.e., $E_t^n$ and $E_t^p$. When generating the plot, we chose $N_1 = N_2 = 40$. One can see that by the end of the time horizon, the sensor used up all the communication opportunities for the perfect channel (inheriting from [19, Fig.6]), but not all the communication opportunities for the noisy channel (inheriting from [25, Fig.3]). This surprising result is due to the fact that if the sensor decides not to transmit its observation, the decoder will know that the realization of the source belongs to a certain interval. This information, which we call "thresholding information", can be more informative than a noisy output from the communication channel plus the side information. Hence, depending on the source realization, the sensor may or may not choose to transmit, even if it is allowed to. More interpretations of a similar result can be found in [25, Remark 9].

## VI. CONCLUSIONS

In this paper, we have analyzed the impact of an additional noisy communication channel over the classical remote estimation problems with one perfect but more costly channel. We have shown that while the intuitive solution of applying threshold-in-threshold communication scheduling policy may be suboptimal for the original problem with soft constraint, it will be optimal, under some mild assumptions, for the setting with a side channel. We have evaluated the performance of optimal decision policies numerically for the problem with hard constraint. The numerical results exhibit several interesting properties that are shared with both the setting with one perfect channel and that with one additive noisy channel.

There are several interesting directions for future research. One is to consider a more general setting with two additive noise channels, where one channel is cheap but noisy, and the other one is costly but less noisy, and in addition, the sensor still has the option of not transmitting its observation. Here, if the sensor decides to transmit its observation, it always sends the observation to the encoder. The encoder generates an encoded message, but may send it to the noisy channel or the less noisy channel, depending on the sensor's decision. The encoder still sends the sign of observation to the decoder via a noiseless side channel. Under assumptions similar to those employed in Theorem 2, it can be shown that the communication scheduling policy can be restricted to "threshold-in-threshold-in-threshold" ones without any loss of optimality. Then, the question that remains to be answered is whether threshold-in-threshold policy, which is a special class within the threshold-in-threshold-in-threshold class, is optimal or not. The analysis in this case seems to be quite complicated, requiring substantial additional effort. Other possible directions for research include consideration of the setting with Markov source instead of i.i.d. source, or the setting with multiple sensors. Yet another direction of research would be to study the effect of adversarial intervention (such as jamming or capture of some of the sensors by an adversary).

## REFERENCES

[1] X. Gao, E. Akyol, and T. Başar, "On remote estimation with multiple communication channels," in *Proceedings of 2016 American Control Conference*, 2016, pp. 5425–5430.

[2] M. Athans, "On the determination of optimal costly measurement strategies for linear stochastic systems," *Automatica*, vol. 8, pp. 397–412, 1972.

[3] K. J. Åström and B. M. Bernhardsson, "Comparison of Riemann and Lebesgue sampling for first order stochastic systems," in *Proceedings of the 41st IEEE Conference on Decision and Control*, 2002, pp. 2011–2016.

[4] Y. Xu and J. Hespanha, "Optimal communication logics for networked control systems," in *Proceedings of the 43rd IEEE Conference on Decision and Control*, 2004, pp. 3527–3532.

[5] V. Gupta, T. H. Chung, B.Hassibi, and R. M. Murray, "On a stochastic sensor selection algorithm with applications in sensor scheduling and sensor coverage," *Automatica*, vol. 42, pp. 251–260, 2006.

[6] M. Rabi, G. V. Moustakides, and J. S. Baras, "Multiple sampling for real-time estimation on a finite horizon," in *Proceedings of the 45th IEEE Conference on Decision and Control*, 2006, pp. 1351–1357.

[7] P. Bommannavar and T. Başar, "Optimal estimation over channels with limits on usage," in *17th IFAC World Congress*, 2008, pp. 6632–6637.

[8] C. Yang and L. Shi, "Deterministic sensor data scheduling under limited communication resource," *IEEE Transactions on Signal Processing*, vol. 59, no. 10, pp. 5050–5056, 2011.

[9] Y. Mo, E. Garone, A. Casavola, and B. Sinopoli, "Stochastic sensor scheduling for energy constrained estimation in multi-hop wireless sensor networks," *IEEE Transactions on Automatic Control*, vol. 56, no. 10, pp. 2489–2495, 2011.

[10] L. Shi and H. Zhang, "Scheduling two Gauss-Markov systems: An optimal solution for remote state estimation under bandwidth constraint," *IEEE Transactions on Signal Processing*, vol. 60, no. 4, pp. 2038–2042, 2012.

[11] Z. Ren, P. Cheng, J. Chen, L. Shi, and Y. Sun, "Optimal periodic sensor schedule for steady-state estimation under average transmission energy constraint," *IEEE Transactions on Automatic Control*, vol. 58, no. 12, pp. 3265–3271, 2013.

[12] K. Nar and T. Başar, "Sampling multidimensional Wiener processes," in *Proceedings of the 53rd IEEE Conference on Decision and Control*, 2013, pp. 3426–3431.

[13] J. Wu, Q. Jia, K. H. Johansson, and L. Shi, "Event-based sensor data scheduling: Trade-off between communication rate and estimation quality," *IEEE Transactions on Automatic Control*, vol. 58, no. 4, pp. 1041–1046, 2013.

[14] K. You and L. Xie, "Kalman filtering with scheduled measurements," *IEEE Transactions on Signal Processing*, vol. 61, no. 6, pp. 1520–1530, 2013.

[15] D. Han, Y. Mo, J. Wu, S. Weerakkody, B. Sinopoli, and L. Shi, "Stochastic event-triggered sensor schedule for remote state estimation," *IEEE Transactions on Automatic Control*, vol. 60, no. 10, pp. 2661–2675, 2015.

[16] S. Weerakkody, Y. Mo, B. Sinopoli, D. Han, and L. Shi, "Multi-sensor scheduling for state estimation with event-based stochastic triggers," *IFAC Proceedings Volumes*, vol. 46, no. 27, pp. 15–22, 2013.

[17] D. Shi, R. J. Elliott, and T. Chen, "Event-based state estimation of discrete-state hidden Markov models," *Automatica*, vol. 65, pp. 12–26, 2016.

[18] M. M. Vasconcelos, A. Nayyar, and U. Mitra, "Optimal sensor scheduling strategies in networked estimation," in *Proceedings of the 56th IEEE Conference on Decision and Control*, 2017, pp. 5378–5384.

[19] O. C. Imer and T. Başar, "Optimal estimation with limited measurements," *International Journal of Systems Control and Communications*, vol. 2, no. 1-3, pp. 5–29, 2010.

[20] G. M. Lipsa and N. C. Martins, "Remote state estimation with communication costs for first-order LTI systems," *IEEE Transactions on Automatic Control*, vol. 56, no. 9, pp. 2013–2025, 2011.

[21] A. Nayyar, T. Başar, D. Teneketzis, and V. V. Veeravalli, "Optimal stategies for communication and remote estimation with an energy harvesting sensor," *IEEE Transactions on Automatic Control*, vol. 58, no. 9, pp. 2246–2260, 2013.

[22] J. Chakravorty and A. Mahajan, "Remote-state estimation with packet drop," *IFAC-PapersOnLine*, vol. 49, no. 22, pp. 7–12, 2016.

[23] ——, "Structure of optimal strategies for remote estimation over Gilbert-Elliott channel with feedback," in *Proceedings of 2017 IEEE International Symposium on Information Theory (ISIT)*, 2017, pp. 1272–1276.

[24] Y. Sun, Y. Polyanskiy, and E. Uysal-Biyikoglu, "Remote estimation of the Wiener process over a channel with random delay," in *Proceedings of 2017 IEEE International Symposium on Information Theory (ISIT)*, 2017, pp. 321–325.

[25] X. Gao, E. Akyol, and T. Başar, "Optimal communication scheduling and remote estimation over an additive noise channel," *Automatica*, vol. 88, pp. 57 – 69, 2018.

[26] E. Akyol, K. Viswanatha, K. Rose, and T. Ramstad, "On zero delay source-channel coding," *IEEE Transactions on Information Theory*, vol. 60, no. 12, pp. 7473–7489, 2014.

[27] E. Akyol, K. Rose, and T. Başar, "Optimal zero-delay jamming over an additive noise channel," *IEEE Transactions on Information Theory*, vol. 61, no. 8, pp. 4331–4344, 2015.

[28] K. Ding, S. Dey, D. E. Quevedo, and L. Shi, "Stochastic game in remote estimation under DoS attacks," *IEEE Control Systems Letters*, vol. 1, pp. 146–151, 2017.

[29] L. Peng, L. Shi, X. Cao, and C. Sun, "Optimal attack energy allocation against remote state estimation," *IEEE Transactions on Automatic Control*, 2017.

[30] X. Gao, E. Akyol, and T. Başar, "On remote estimation with communication scheduling and power allocation," in *Proceedings of the 55th IEEE Conference on Decision and Control*, 2016, pp. 5900–5905.

[31] ——, "Optimal sensor scheduling and remote estimation over an additive noise channel," in *Proceedings of 2015 American Control Conference*, 2015, pp. 2723–2728.

[32] ——, "Optimal estimation with limited measurements and noisy communication," in *Proceedings of the 54th IEEE Conference on Decision and Control*, 2015, pp. 1775 – 1780.

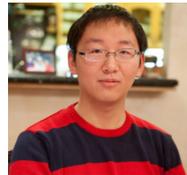

**Xiaobin Gao** received his B.S. degree in Electrical Engineering from the University of Michigan, Ann-Arbor, in 2012, and another B.S. degree in Mechanical Engineering from Shanghai Jiao Tong University in the same year. He received his M.S. degree in Electrical Engineering from the University of Illinois, Urbana-Champaign, in 2014. He is pursuing his Ph.D. degree now in Electrical and Computer Engineering at the University of Illinois, Urbana-Champaign, where he is a research assistant at the Coordinated Science Laboratory. His research interest includes sensor networks, remote estimation, cyber-physical systems, switched systems, and distributed computation.

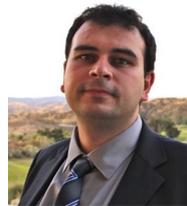

**Emrah Akyol** is an Assistant Professor of Electrical and Computer Engineering at the State University of New York at Binghamton. He received his Ph.D. degree in 2011 from the University of California at Santa Barbara. From 2006 to 2007, he held positions at Hewlett-Packard Laboratories and NTT Docomo Laboratories, both in Palo Alto, CA where he worked on topics in image and video compression. From 2013 to 2014, Dr. Akyol was a postdoctoral researcher in the Electrical Engineering Department at University of Southern California, between 2014 and 2017, in the Coordinated Science Laboratory at University of Illinois at Urbana-Champaign. His current research is on the applications of game, communication and control theories and machine learning to the design and analysis of secure cyber-physical systems. He is a senior member of IEEE.

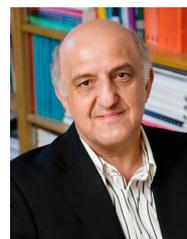

**Tamer Başar** (S'71-M'73-SM'79-F'83-LF'13) is with the University of Illinois at Urbana-Champaign, where he holds the academic positions of Swanlund Endowed Chair; Center for Advanced Study Professor of Electrical and Computer Engineering; Research Professor at the Coordinated Science Laboratory; and Research Professor at the Information Trust Institute. He is also the Director of the Center for Advanced Study. He received B.S.E.E. from Robert College, Istanbul, and M.S., M.Phil, and Ph.D. from Yale University. He is a member of the US National Academy of Engineering, member of the European Academy of Sciences, and Fellow of IEEE, IFAC (International Federation of Automatic Control) and SIAM (Society for Industrial and Applied Mathematics), and has served as president of IEEE CSS (Control Systems Society), ISDG (International Society of Dynamic Games), and AACC (American Automatic Control Council). He has received several awards and recognitions over the years, including the highest awards of IEEE CSS, IFAC, AACC, and ISDG, the IEEE Control Systems Award, and a number of international honorary doctorates and professorships. He has over 800 publications in systems, control, communications, networks, and dynamic games, including books on non-cooperative dynamic game theory, robust control, network security, wireless and communication networks, and stochastic networked control. He was the Editor-in-Chief of Automatica between 2004 and 2014, and is currently editor of several book series. His current research interests include stochastic teams, games, and networks; security; and cyber-physical systems.